\documentclass[aps,prl,superscriptaddress,showpacs,twocolumn]{revtex4}

\usepackage{epsfig}
\usepackage{subfigure}
\usepackage{times}
\usepackage{amsmath}
\usepackage{amsthm}
\usepackage{bm,amssymb,verbatim}
\usepackage{amsbsy}

\newcommand{\expect}[1]{\langle {#1} \rangle}

\begin{document}
\title{Quench induced Mott insulator to superfluid quantum 
phase transition}

\author{Jay D.~Sau}
\affiliation{Condensed Matter Theory Center and Joint Quantum Institute, Department of Physics,
University of Maryland, College Park, Maryland 20742-4111, USA}
\author{Bin Wang}
\affiliation{Condensed Matter Theory Center and Joint Quantum Institute, Department of Physics,
University of Maryland, College Park, Maryland 20742-4111, USA}
\author{S. Das Sarma}
\affiliation{Condensed Matter Theory Center and Joint Quantum Institute, Department of Physics,
University of Maryland, College Park, Maryland 20742-4111, USA}

\begin{abstract}
 Mott insulator to superfluid quenches have been used by recent experiments
 to generate exotic superfluid phases. While the final Hamiltonian
 following the sudden quench is that of a superfluid, it is not
 apriori clear how close the final state of the system approaches
 the ground state of the superfluid Hamiltonian. To understand the
 nature of the final state
 we calculate the temporal evolution of the momentum distribution
 following a Mott insulator to superfluid quench.
Using the numerical infinite time-evolving block decimation
approach and the analytical rotor model approximation we establish that the
one and two dimensional Mott insulators following the quench
equilibriate to thermal states with spatially short-ranged
coherence peaks in the final momentum distribution and 
therefore are not strict superfluids. However, in
three dimensions we find a divergence in the momentum distribution
indicating the emergence of true superfluid order.
\end{abstract}
\pacs{03.75.Kk,05.30.Rt}

\maketitle

\paragraph{Introduction}
\label{Sec:Intro} Systems of ultra-cold atoms provide us with
many-body systems with tunable Hamiltonians that may potentially
be used to realize various quantum phases of matter. However many 
 of the interesting phases that have been predicted require 
extremely low temperatures
that  have not been reached in experiments.
 One new phase of matter that has been
realized in cold atomic systems is the Mott-insulator (MI) phase,
 which is a gapped collective phase and therefore has the special
 property of a low entropy-density.
 This low entropy density makes the MI a favorable starting point for
 creating other low-temperature
 phases of atoms. In particular, a recent set of
 experiments \cite{Hemmerich} uses an effective
Mott phase as a starting state and suddenly ramps up the tunneling
to create potentially exotic low temperature superfluid-like
 (SF) states with $p$ and $f$-orbital 
 symmetries that have never been realized before.
While such a sudden change or quench of the Hamiltonian from an
initially known state to a desired Hamiltonian is a promising way
to realize new states of matter  \cite{stamperkurn,liu}, it is not
clear whether a system following the quench evolves to a state
near the desired ground state phase in any sense.
Since the quenching process is manifestly non-adiabatic and very 
far from equilibrium, it is certainly possible, perhaps even likely, 
that the final state has only a small overlap with the ground state 
of the final Hamiltonian!
Measurements of the momentum density using time-of-flight imaging
do indeed show that the initially uniform momentum distribution,
which characterizes the MI phase, develops coherence peaks that
are characteristic of the superfluid phase. However, this
evolution where the momentum density bunches up into peaks instead
of spreading out is counter-intuitive and suggests dissipation of
energy similar to evaporative cooling that is used to create
Bose-Einstein condensates (BECs). Therefore, to understand the
utility of this approach to realizing
 new phases of ultra-cold atoms,
it is necessary to understand both the origin and extent of the
superfluid state that is formed following similar quenches.

A natural approach to understanding these quenches is to consider
an analogous quench in the most studied and simplest transition
that has been observed in ultra-cold atomic systems, namely the MI
to SF phase transition in $s$-wave optical lattices
\cite{greiner,kasevich}.
In addtion there have been several theoretical studies of quenches 
across the MI-SF transition \cite{altman,polkovnikov,kasevich,sfmiquench,rigol,quench2010} which have focussed on aspects such as the 
short-time coherent dynamics or the development of local phase
 coherence and entanglement in real space which is a particularly
interesting aspect of  MI to SF quenches.
 However, these studies do not completely characterize the final state
 of the system reached. 
In particular they contain only indirect information about the
 long-range coherence that is measured by time-of-flight imaging.
Therefore some very recent works \cite{1Dquench_expt,3Dquench_expt}
 have concentrated on comparing the 
time-of-flight measurements in quench experiments with
 theoretical estimates of the 
momentum density.

In this letter we develop an understanding of the peak formation in
  momentum
distribution  following an MI-SF quench by calculating this
 distribution for 
 bosonic atoms trapped in an $s$-wave optical lattice.
 Since the initial MI phase
 is completely incoherent between sites, the momentum distribution
 is uniform. To reach a SF state,
the tunneling is instantaneously quenched to a finite value on the
SF side of the ground-state phase diagram.
 We find that the final momentum distribution after the quench
 depends strongly on the dimensionality of the lattice under consideration.
The simplest case is the zero-dimensional (0D) lattice, where the
momentum density following the MI-SF quench continues to oscillate
in time and thus does not appear to equilibrate. In contrast, the
one-dimensional (1D) MI chain, whose dynamics we calculate numerically
using the systematically exact infinite time-evolving block
decimation (iTEBD) \cite{itebd} approach yields a momentum
distribution which approaches equilibrium.
Furthermore, the dynamics of the momentum density  qualitatively
agrees with the distribution calculated from the 
quantum rotor model (exact in the large filling limit)
\cite{largeN,rotor} studied within the semi-classical
 truncated Wigner approximation (TWA)\cite{twa}.
Moreover, the equilibrium momentum density estimated from iTEBD 
quantitatively agrees with the result from the classical rotor
model.
 Using this fact, which we verified in 0D and one
dimensions (1D), we estimate the equilibrium momentum density using the
classical rotor approximation in two dimensions (2D) and three dimensions 
(3D)
following a quench, which are relevant to the recent
experiments \cite{Hemmerich}. We find that such an MI-SF quench in
the 2D limit leads to a classical thermal gas state
with short-ranged momentum correlations (and no condensate
fraction). On the other hand, an MI-SF quench in the 3D
 case, which directly applies to the experimental
results, yields a true condensate.

\paragraph{Quenches in the Bose Hubbard model:}
Ultra-cold bosonic atoms in a deep optical lattice can be described by the
 Bose-Hubbard model which is written as
\begin{equation}
H=-\zeta\sum_{\langle i,j\rangle}(b_i^\dagger b_j+h.c)+
\frac{U}{2}\sum_j n_j(n_j-1),
\end{equation}
where $\langle i,j\rangle$ are nearest neighbor sites on the
optical lattice, $b_j^\dagger$ is the boson creation operator on
site $j$, $n_j=b_j^\dagger b_j$ is the number operator of bosons
on site $j$ and $U$ is the charging energy per site. The
inter-site tunneling matrix element $\zeta$ between the sites is
taken to vanish in the initial state so that the system is
initially in a MI ground state with $\langle n_j\rangle=n$
(integer) atoms per site. The momentum distribution of the bosons,
that is measured by time-of-flight images, is the fourier
transform of the off-diagonal density matrix
$n_k=\sum_j e^{i k j}\expect{b^\dagger_j b_0}$.
\begin{figure}
\centering
\includegraphics[scale=0.4,angle=270]{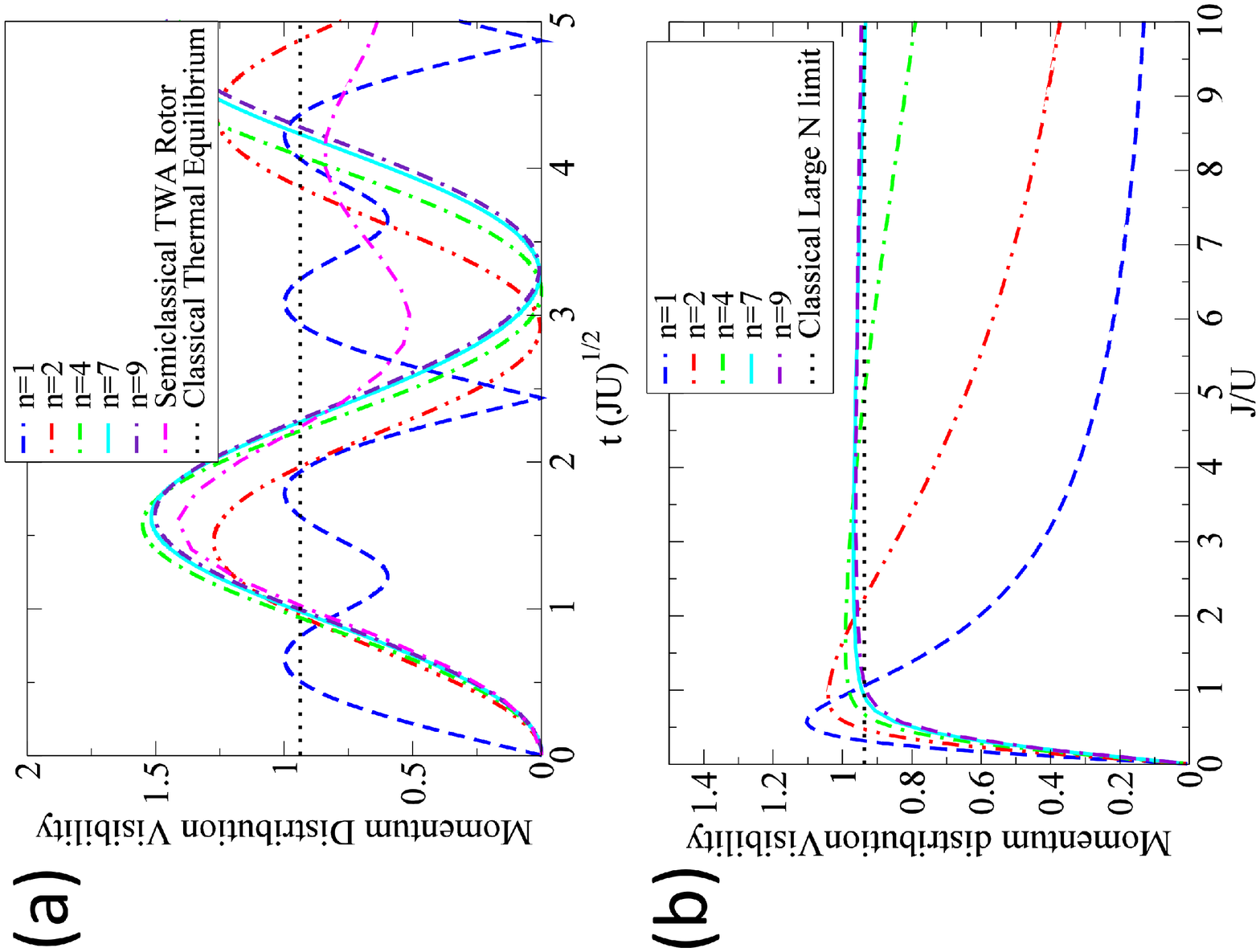}
\caption{(a)Quantum dynamics simulations of visibility for 0D system show
absence of equilibration. 
(b) Equilibrium visibility of momentum distribution (defined in Eq.~\ref{eq:visibility}) following a quench in a 0D (2 site) Mott insulator
from $J_{initial}=0$ to $J_{final}=J$ at various fillings
 $n$ as a function of $J/U$. Equilibrium is assumed to be thermal.
 Results at large filling $n$ agree well with the rotor value.
 }\label{Fig1}
\end{figure}
When the state following the quench reaches a thermal
equilibrium, the correlations can be calculated from the free
energy
\begin{equation}
F[\beta,\mu]=-\frac{1}{\beta}\log{Tr(e^{-\beta H-\mu N})}\label{eq:F}
\end{equation}
where the inverse final temperature $\beta$ and chemical potential $\mu$
are determined by the conserved initial energy $\expect{E}=-\partial_\beta (\beta F)=0$ and $\expect{N}=-\beta\partial_\mu F=n$.

\paragraph{0D quenches:}
Let us start by considering the simplest MI-SF quench of a
0D MI which has two sites connected by a bond forming a two-well Josephson junction\cite{rotor}.
The Hamiltonian $H$ for such a system can be numerically diagonalized
exactly to obtain both the
equilibrium value and dynamics of the momentum distributions $n_k$,
which we plot in Fig.~\ref{Fig1}. For convenience we focus on
the ``visibility'' of the momentum distribution,
 \begin{equation}
\textrm{Visibility}=\frac{2(n(k=0)-n(k=\pi/a))}{(n(k=0)+n(k=\pi/a))},\label{eq:visibility}
\end{equation}
which is a measure of how peaked the momentum distribution is at
its maximum at $k=0$ compared to its minimum at $k=\pi/a$ ($a$
being the lattice constant of the optical lattice).
The visibility following a quench in the
 0D MI-SF quench (Fig.~\ref{Fig1}(a)) oscillates 
as has been found in previous results \cite{altman,polkovnikov} and 
does not equilibrate in contrast to higher dimensional
MIs where we find that $n_k$ reaches the 
equilibrium distribution calculated from Eq.~\ref{eq:F}.
Therefore it is useful to consider the results of the thermal
distribution in Fig.~\ref{Fig1}(b).

Consistent with previous theoretical studies \cite{polkovnikov,rotor},
the results in Fig.~\ref{Fig1} show that for reasonably large filling 
factors $n$ (other than $n=1$), the time-scale for dynamics and maximum 
visibility attain universal values for intermediate $J/U$  corresponding
 to the quantum rotor model:
\begin{equation}
H_{rot}=-J\sum_{\langle ij\rangle}\cos{(\phi_i-\phi_j)}+\frac{U}{2 }\sum_j \pi_j^2\label{eq:hrot}
\end{equation}
where $\pi_j=-i\partial_{\phi_j}$ is the angular momentum
canonically conjugate to $\phi_j$\cite{largeN,rotor} and 
$J=2 n \zeta$ is the Josephson coupling. The
free-energy $F$ is approximated by replacing $H\rightarrow
H_{rot}$ in Eq.~\ref{eq:F}. At values of $J$  comparable to the
Mott-gap $U$, the system equilibrates to a sufficiently high
temperature $\beta^{-1}$, so that the free-energy $F$ in the
classical limit \cite{polkovnikov} is a combination
 ($F=-\frac{\log{Z_{XY}}}{\beta}+\frac{\log{\beta}}{2\beta}$)
 of charge fluctuations with energy $\frac{1}{2\beta}$ and 
phase fluctuations described the classical XY model
\begin{equation}
H_{XY}=-J \sum_{\langle ij\rangle}\cos{(\phi_i-\phi_j)}\label{eq:hxy}
\end{equation}
 where $Z_{XY}=\int \prod_j d\phi_j e^{-\beta H_{XY}}$ is the 
partition function of the $XY$ model.
In this approximation, $\beta^{-1}$ is determined by the energy conservation equation
\begin{equation}
Z_{XY}^{-1}\int \prod_j d\phi_j \cos{(\phi_1-\phi_0)}e^{\tilde{J}\sum_{\langle j,k\rangle}\cos{(\phi_j-\phi_k)}}=\frac{1}{2 d\tilde{J}}\label{eq:Tfinal}
\end{equation}
where $\tilde{J}=\beta J$ and $d$ is the average coordination number
 of each site of the lattice under consideration. For the 0D
lattice, $d=1$ (after dropping a global $U(1)$ degree of freedom).
The equilibrium momentum distribution $n_k$ is given by the Fourier
 transform of the real-space correlation function 
\begin{equation}
\langle b_j^\dagger b_0\rangle\propto\langle e^{i (\phi_j-\phi_0)}\rangle=\int \frac{\prod_l d\phi_l}{Z_{XY}} e^{i (\phi_j-\phi_0)} e^{\tilde{J}\sum_{\langle ll'\rangle}\cos{(\phi_{l}-\phi_{l'})}}.
\end{equation}
Within the $XY$-model approximation, the coherent momentum distribution
 peak results from a dissipation of phase fluctuation energy $H_{XY}$
in the initial state into charge fluctuation energy in the final state.
The 0D MI equilibrium is given by
 $\tilde{J}=1.065$ so that one has an equilibrium
visibility of $=0.939$ which is in good agreement with the large $n$ 
equilibrium in Fig.~\ref{Fig1}(b).
At large values of $J$, fluctuations in the number density
 ($|(n_j-n)=\pi_j|\gtrsim n$) invalidate the simple rotor approximation $H_{rot}$.

\begin{figure}
\centering
\includegraphics[scale=0.4,angle=270]{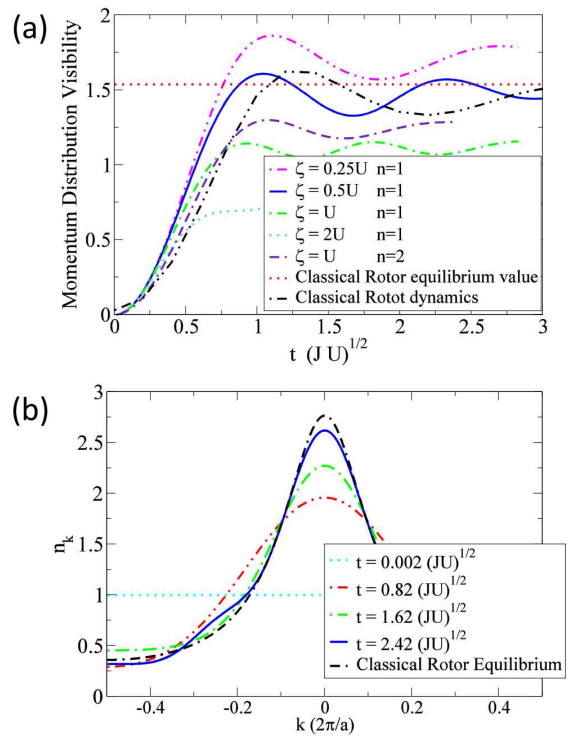}
\caption{(a) Quantum dynamics (iTEBD) simulations of visibility
 for 1D system  and TWA show equilibration following a quench in a MI 
from $J_{initial}=0$ to $J_{final}=J$ at various fillings
 $n$ and $J/U$. The thermal equilibrium visibility determined from the
XY model is in good agreement with the iTEBD and TWA results.
(b) Time-evolution of the momentum distribution for $J=0.5\,U$ and $n=1$
shows evolution of the momentum distribution to one that is in
good agreement with the classical thermal distribution. }\label{Fig2}
\end{figure}

\paragraph{1D Mott insulators:}
Unlike the 0D MI, the visibility 
peak of the 1D MI (Fig.~\ref{Fig2}(b)), calculated 
using the in principle exact iTEBD
method, appears to tend to a steady state value.
 These results agree qualitatively with 
semiclassical dynamics associated with the quantum rotor model
$H_{rot}$ calculated within the TWA \cite{polkovnikov}.
 Within the TWA, the distribution of the phases $\phi_j$
 begins with the uniform
distribution for the MI phase and evolves according to the
classical equations of motion \cite{polkovnikov,twa}
\begin{equation}
\ddot{\phi}_j=J U \sum_{i}\sin{(\phi_j-\phi_i)}\label{eq:eom}.
\end{equation}
Thus, as seen in Fig.~\ref{Fig2}(b), the time-scale for dynamics
following the MI-SF quench within iTEBD and also the classical
rotor approximation scales as  $\tau\propto (J U)^{-1/2}$.
However, as in the 0D case, the classical 
estimate for the final equilibration visibility
 appears to be valid only for intermediate values of 
 $J/U$.

The partition function for the classical 0D XY model
 (Eq.~\ref{eq:hxy}), can be written in a transfer-matrix form
 (see for example \cite{transfermatrix_lg}) as
$Z_{XY}=Tr[M^N(\phi-\phi')]$ where
 $M(\phi)=e^{\tilde{J}\cos{\phi}}=\sum_n I_n(\tilde{J})e^{i n \phi}$, 
and $I_n(x)$ are modified Bessel functions.
 The equilibrium momentum distribution of the bosons for the 
$N=\infty$ lattice is calculated to be
\begin{equation}
n_k=\sum_r e^{i k r}\langle e^{i (\phi_r-\phi_0)}\rangle=\frac{4\tilde{J}^2-1}{4\tilde{J}^2+1-4\tilde{J}\cos{k}}.
\end{equation}
The classical equilibrium visibility from Eq.~\ref{eq:visibility} 
at the equilibrium parameter $\tilde{J}=1.065$, consistent with
 Fig.~\ref{Fig2}(a), is found to be 1.52.
Moreover, the final momentum distribution at $n=1$, $\zeta=0.5 U$ shown in Fig.~\ref{Fig2}(b) is in quantitative agreement
with the classical equilibrium $n_k$.
\begin{figure}
\centering
\includegraphics[scale=0.3,angle=270]{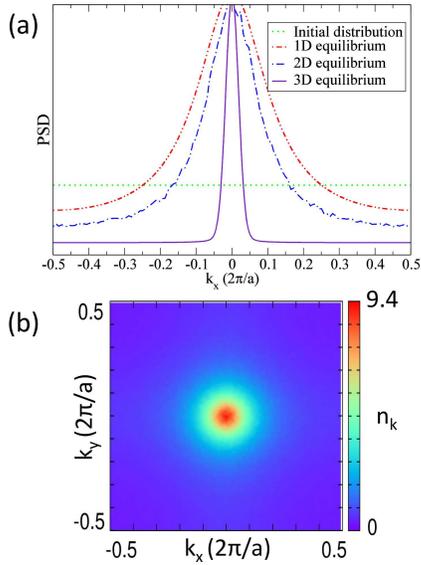}
\caption{(a) Equilibrium momentum density as a function of momentum $k_x$
 after an
MI-SF quench as compared with the initial distribution
for 0D, 2D and 3D MIs. All calculated within the classical rotor model
 approximation. The 3D MI case shows the emergence
 of a true condensate with a divergent $k=0$ peak which we have
 broadened by $\sigma=0.05\pi/a$. The momentum densities
 at $k_x=0$ has been scaled to the same value.
(b) 2D momentum distribution $(n_{\bm{k}})$ following
 an MI-SF quench for the 2D MI shows a peak at
 $k_x=k_y=0$ which is 3 times higher than the 1D peak
and qualitatively resembles the peak seen
in experiments \cite{Hemmerich}. }\label{Fig3}
\end{figure}

\paragraph{2D Mott insulators:}
The classical rotor model approximation can also be used to calculate
 the equilibrium momentum distribution for a MI-SF quench in a 2D
 MI at filling
factors $n\gg 1$ in the initial state. Such a
2D MI quench could be realized in experiments by
adding a planar confining potential to the set-up in recent
experiments \cite{Hemmerich}.
Here we use the 
cluster-update Monte-Carlo (MC) scheme \cite{wolff} to calculate
the average energy for the 2D XY model 
(Eq.~\ref{eq:hxy}) on a $127\times 127$ lattice with periodic
boundary conditions to determine the equilibrium final temperature
satisfying Eq.~\ref{eq:Tfinal}. We find the equilibrium
temperature $\beta^{-1}=1.51 J$ to be larger than the
Kosterlitz-Thouless temperature $\beta^{-1}_{KT}=0.89 J$
\cite{KT}.
 As seen in Fig.~\ref{Fig3}, the momentum density $n_k$ at
 $\tilde{J}=\beta J=0.662$ calculated using MC shows a peak (Fig.~\ref{Fig3}) at
 $k=0$ which is 3 times higher than the 1D case and is qualitatively similar to the momentum distributions observed
 in experiment for $p-$wave superfluids \cite{Hemmerich}.
 However, the equilibrium momentum distributions of the
quenched MI both in 1D and 2D remain finite and
smooth around $k\sim 0$ indicating a boson correlator that is
exponentially decaying in real space.

\paragraph{3D Mott Insulator:}The classical rotor approximation
can also be applied to the 3D case as well. In
contrast to the lower dimensional MI-SF quenches, we find that in
3D, the energy density $\epsilon_{ns}=-0.988$ at
MI-SF transition calculated using MC \cite{0202017} is slightly
higher than the equilibrium value following the quench given by
Eq.~\ref{eq:Tfinal}, indicating a equilibrium phase with a true SF
component. This is consistent with the high-temperature expansion
for a 3D MI on a simple-cubic lattice \cite{3DhighT}  which
estimates the equilibrium temperature $\beta^{-1}\sim 2.10 J$ to
be below the critical temperature $T_c=2.202 J$ \cite{0202017}.
 Since the equilibrium temperature is close to
the critical temperature, we can use scaling relations around the
critical point to estimate the momentum distribution and the
equilibrium temperature below $T_c$.
 Using the scaling relations for the energy density \cite{critical_review}
around $\beta^{-1}\sim T_c$, we obtain the equilibrium temperature
$\beta^{-1}=T_c-0.077 J$. The momentum distribution in the
vicinity of $\beta^{-1}\lesssim T_c$ has a universal form at small
$k$ \cite{halperin,critical_review} which is given by
\begin{equation}
F(k)\approx  w(\beta)^2[(2\pi)^3\delta(\bm k)+\frac{\xi_T}{k^2}]+f_0.\label{eq:3Dn}
\end{equation}
 Since the critical fluctuation
contribution to $n_k$ is only accurate near $k\sim 0$, we have
added a $k$-independent background $f_0=[1-w(T)^2(1+\xi_T/2\pi)]$
to restore the total spectral weight to unity. Here $w(\beta)=B
\left(\frac{T_c-\beta^{-1}}{T_c}\right)^{\beta_0}$ is the
condensate fraction of the system. The parameters in
Eq.~\ref{eq:3Dn} have been calculated to be:
 the critical amplitude $B=1.245$ , the
critical exponent $\beta_0\approx 0.35$, the transverse
correlation length
$\xi_T(T)=\xi^{(-)}\left(\frac{T_c-T}{T_c}\right)^{-\nu}$ with
$\xi^{(-)}_T\approx 1.65$ and $\nu\approx 0.67$
\cite{critical_amplitude,halperin,3DhighT,0202017,critical_review}. To resolve the singularity at $k\sim 0$ 
we have broadened the momentum distribution by a 
width $\sigma=0.05\pi/a$. The resulting singular momentum distribution
 is plotted in Fig.~\ref{Fig3} as a function of $k_x=k$, with $k_y=k_z=0$ 
has a peak near $k\sim 0$ with total weight of $1-f_0\sim 0.3$.

The recent MI-SF quench experiments \cite{Hemmerich}, in the
3D limit, can be thought of as a 3D MI-SF quench
where the initial state is the ground state of the 3D XY
Hamiltonian (Eq.~\ref{eq:hxy}) with tunneling $\zeta$ turned on
only along $z$. This corresponds to a lower initial energy density
which in turn leads to a reduced equilibrium temperature
$T_c-\beta^{-1}=0.16 J$ because of energy dissipated in the 
phase-fluctuations along $z$.
 This results peak near $k\sim 0$ in the momentum distribution 
with
total weight $1-f_0\approx 0.6$ which is twice as large as 
the peak obtained from quenching from the 3D MI state.

\paragraph{Conclusion:} 
We have calculated the formation of peaks in the momentum
distribution of bosonic atoms in an optical lattice following a
quench from deep in the MI phase to an SF phase within the
classical XY model approximation. The classical
results are in semi-quantitative agreement with exact iTEBD
calculations for the 1D MI. Understanding such MI-SF
quenches is particularly important in the context of recent
experiments that use such quenches to reach superfluid-like
states. From our calculations we find that in 1D, 2D and
3D MIs, the initially uniform momentum distribution
develops a peak at momentum $k=0$ following the quench that is
qualitatively similar to the one in experimental time-of-flight
images. However in 1D and 2D the resulting momentum
distributions are non-singular at $k=0$ and represent phases without
 long-range phase order. In contrast, we find a true condensate
 in the 3D MI
corresponding to the recent experiments\cite{Hemmerich} with a
divergent momentum distribution at $k=0$.
One interesting way to verify our theoretical predictions is to 
use the recently developed quantum gas microscope technique
 \cite{greiner1} to look directly at the time evolution of the
 MI to SF quenches in systems of different dimensionalities.

We thank R. Sensarma and A. Hemmerich for introducing us to this
problem. This work was supported by  DARPA-OLE, JQI-NSF-PFC, ARO-MURI 
and AFOSR-MURI.


\begin{thebibliography}{35}
\expandafter\ifx\csname natexlab\endcsname\relax\def\natexlab#1{#1}\fi
\expandafter\ifx\csname bibnamefont\endcsname\relax
  \def\bibnamefont#1{#1}\fi
\expandafter\ifx\csname bibfnamefont\endcsname\relax
  \def\bibfnamefont#1{#1}\fi
\expandafter\ifx\csname citenamefont\endcsname\relax
  \def\citenamefont#1{#1}\fi
\expandafter\ifx\csname url\endcsname\relax
  \def\url#1{\texttt{#1}}\fi
\expandafter\ifx\csname urlprefix\endcsname\relax\def\urlprefix{URL }\fi
\providecommand{\bibinfo}[2]{#2}
\providecommand{\eprint}[2][]{\url{#2}}


\bibitem{Hemmerich} G. Wirth, M. \"{O}lschl\"{a}ger, and A. Hemmerich,  Nat. Phys.\textbf{7}, 147 (2011); M. \"{O}lschl\"{a}ger, G. Wirth, and A. Hemmerich, Phys. Rev. Lett. \textbf{106} 015302 (2011).

\bibitem{stamperkurn}L. E. Sadler, J. M. Higbie, S. R. Leslie, M. Vengalattore, and D.M. Stamper-Kurn, Nature \textbf{443}, 312 (2006).

\bibitem{liu}M. Lewenstein and W. V. Liu, Nat. Phys.,\textbf{7}, 101, (2011).

\bibitem{greiner}M. Greiner et al, Nature \textbf{415}, 39 (2002).


\bibitem{kasevich}  C. Orzel, A. K. Tuchman, M. L. Fenselau, M. Yasuda and
 M. A. Kasevich, Science \textbf{291}, 2386 (2001);
A. K. Tuchman, C. Orzel, A. Polkovnikov, and M. A. Kasevich,  Phys. Rev. A \textbf{74}, 051601(R) (2006).



\bibitem{altman}E. Altman, A. Auerbach, Phys. Rev. Lett. \textbf{89}, 250404 (2002).

\bibitem{polkovnikov}A. Polkovnikov, S. Sachdev and S. M. Girvin, Phys. Rev. A 66, 053607 (2002)



\bibitem{sfmiquench} C. Kollath, A. Lauchli and E. Altman, Phys. Rev. Lett. \textbf{98}, 180601 (2007); G. Roux, Phys. Rev. A, \textbf{79}, 021608 (2009).

\bibitem{rigol} M. Rigol, A. Muramatsu, Phys. Rev. Lett. \textbf{94}, 240403 (2005).

\bibitem{quench2010} P. Navez, R. Schutzhold, Phys. Rev. A \textbf{82}, 063603 (2010).

\bibitem{1Dquench_expt}S. Trotzky, Y.-A. Chen , A. Flesch, I. P. McCulloch, U. Schollwock1, J. Eisert and I. Bloch, 
arXiv:1101.2659 (2011).

\bibitem{3Dquench_expt}D. Chen, M. White, C. Borries, and B. DeMarco, arXiv:1103.4662 (2011).

\bibitem{itebd} G. Vidal, Phys. Rev. Lett. \textbf{98}, 070201 (2007).

\bibitem{largeN} U.R. Fischer, R. Schutzhold, M. Uhlmann, Phys. Rev. A \textbf{77}, 043615 (2008).



\bibitem{rotor}J. R. Anglin, P. Drummond, and A. Smerzi, Phys. Rev. A \textbf{64},
063605 (2001);R. Barnett, J. D. Sau, and S. Das Sarma, Phys. Rev. A \textbf{82}, 031602(R) (2010).



\bibitem{twa}J. D. Sau, S. R. Leslie, D. M. Stamper-Kurn, and M. L. Cohen, Phys. Rev. A \textbf{80}, 023622 (2009); A. Polkovnikov, Annals of Phys. \textbf{325}, 1790 (2010)

\bibitem{transfermatrix_lg}D.J. Scalapino, M. Sears, R. A. Ferrell, Phys. Rev. B,
\textbf{6},3409 (1972).

\bibitem{wolff}U. Wolff, Phys. Rev. Lett, \textbf{62}, 361 (1989).

\bibitem{KT}J.M. Kosterlitz and D. J. Thouless, J. Phys. C \textbf{6}, 1181 (1973);
J. Tobochnik and G. V. Chester, Phys. Rev. B \textbf{20}, 3761 (1979).


\bibitem{3DhighT} M. Ferer, M. A. Moore, M. Wortis, Phys. Rev. B, \textbf{8}, 5205 (1973).

\bibitem{critical_amplitude} A. Aharony and P. C. Hohenberg, Phys. Rev. B \textbf{13}, 3081 (1976).
\bibitem{halperin}P. C. Hohenberg, A. Aharony, B. I. Halperin and E. D. Siggia, Phys. Rev. B \textbf{13}, 3081 (1976).

\bibitem{0202017}A. Cucchieri, J. Engels, S. Holtmann, T. Mendes, T. Schulze, J. Phys. A, \textbf{35},6517 (2002).

\bibitem{critical_review}A. Pelissetto and E. Vicari, Phys.Rept. \textbf{368},549 (2002).
\bibitem{greiner1}W. S. Bakr, J. I. Gillen, A. Peng, S. Fölling,  and  M. Greiner, Nature \textbf{462}, 74 (2009) 
\end{thebibliography}
\end{document}